\documentclass[conference]{IEEEtran}
\IEEEoverridecommandlockouts
\usepackage{cite}
\usepackage{amsmath,amssymb,amsfonts}
\usepackage{algorithmic}
\usepackage{graphicx}
\usepackage{subcaption}
\usepackage{multirow}
\usepackage{textcomp}
\usepackage{xcolor}
\def\BibTeX{{\rm B\kern-.05em{\sc i\kern-.025em b}\kern-.08em
    T\kern-.1667em\lower.7ex\hbox{E}\kern-.125emX}}

\begin{document}

\title{A Study on Robustness to Perturbations for Representations of Environmental Sound\\
\thanks{This work is partially supported by the NSF award 1544753.}
}


\author{\IEEEauthorblockN{Sangeeta Srivastava$^{1}$, Ho-Hsiang Wu$^{2}$, Joao Rulff$^{2}$, Magdalena Fuentes$^{2}$, \\
Mark Cartwright$^{3}$,
Claudio Silva$^{2}$, Anish Arora$^{1}$, Juan Pablo Bello$^{2}$}
\\
$^1$The Ohio State University, Columbus, OH, USA\\
\IEEEauthorblockA{$^2$New York University, New York, NY, USA\\
$^3$New Jersey Institute of Technology, New Jersey, NY, USA\\
}}

\maketitle

\begin{abstract}
Audio applications involving environmental sound analysis increasingly use general-purpose audio representations, also known as embeddings, for transfer learning. Recently, Holistic Evaluation of Audio Representations (HEAR) evaluated twenty-nine embedding models on nineteen diverse tasks. However, the evaluation's effectiveness depends on the variation already captured within a given dataset. Therefore, for a given data domain, it is unclear how the representations would be affected by the variations caused by myriad microphones' range and acoustic conditions -- commonly known as \textit{channel effects}. We aim to extend HEAR to evaluate invariance to channel effects in this work. To accomplish this, we imitate channel effects by injecting perturbations to the audio signal and measure the \textit{shift} in the new (perturbed) embeddings with three distance measures, making the evaluation domain-dependent but not task-dependent. Combined with the downstream performance, it helps us make a more informed prediction of how robust the embeddings are to the channel effects. We evaluate two embeddings -- YAMNet, and OpenL$^3$ on monophonic (UrbanSound8K) and polyphonic (SONYC-UST) urban datasets. We show that one distance measure does not suffice in such task-independent evaluation. Although Fr\'echet Audio Distance (FAD) correlates with the trend of the performance drop in the downstream task most accurately, we show that we need to study FAD in conjunction with the other distances to get a clear understanding of the overall effect of the perturbation. In terms of the embedding performance, we find OpenL$^3$ to be more robust than YAMNet, which aligns with the HEAR evaluation.


\end{abstract}

\section{Introduction}
The scarcity of a large amount of labeled data for supervised learning in applications related to environmental sounds has popularized the use of representation learning and transfer learning \cite{cramer2019look, jansen2020coincidence, wang2021multi, srivastava2021conformer, alayrac2020self} in such applications. As part of this learning paradigm, a network is pre-trained on an \textit{upstream} task, which has the availability of large datasets to learn generic representations or embeddings that are transferable across a variety of related target \textit{downstream} application(s). Bengio et al. \cite{bengio2013representation} defines good representations as one that are \textit{expressive} enough to capture a considerable number of possible input configurations, are \textit{invariant} to most local changes of the input, and \textit{disentangles} the factors of variation in the input.

With the increase in the number of learning frameworks and architectures to learn such invariant representations, there is a need for evaluation benchmarks to test the generalization of the embedding models and empirically compare them. Holistic Audio Representation Evaluation Suite (HARES) \cite{wang2021towards} and Holistic Evaluation of Audio Representations (HEAR) \cite{turian2022hear} are two efforts at this front that test the invariance of various audio representations to downstream domains and tasks. The HEAR challenge, in particular, is the most extensive effort to date and includes an evaluation of twenty-nine audio embedding models on nineteen diverse downstream tasks. However, both HEAR and HARES evaluation methodologies have several limitations. Firstly, they are dependent on the included tasks, as well as the quantity and distribution of the training and test sets of those tasks. Thus, they are not informative as to how the embeddings will perform on unseen tasks. Secondly, they do not provide information on the stability of the audio embeddings in response to specific changes in the same data domain. Hence, not only might they give a limited understanding of what to expect when employing them in real-world applications under various conditions, but they also require inspection and analysis of their test sets to gain understanding of their stability. Lastly, they rely on the availability of annotated data for evaluation. This has the inherent drawback of requiring human annotations, especially for data collected from real-world deployments \cite{bello2019sonyc, catlett2017array}.

In this work, we propose a path to address these limitations by using the following two steps in the evaluation framework: \textbf{(i)} we propose an alternative, yet complementary, testing scenario that includes \textit{invariance to channel effects}. To accomplish this, we artificially degrade audio signals \cite{de2021analysis, serizel2020sound, salamon2017scaper} by applying different mathematical transformations or \textit{perturbations}, and \textbf{(ii)} we leverage distance metrics that quantify the \textit{shift} in the embedding space directly, making the evaluation independent of the task but still dependent on the data domain. We correlate the metrics with the downstream results to corroborate the findings and establish the relationship between the perturbations and the downstream evaluation. We leverage two publicly available audio embedding models, \textit{OpenL$^3$} \cite{cramer2019look} and \textit{YAMNet\footnote{https://github.com/tensorflow/models/tree/master/research/audioset/yamnet}}, to build on the findings from the HEAR challenge, where OpenL$^3$ and YAMNet are among the best and worst-performing models, respectively.

Our contributions can be summarized as follows:
\begin{enumerate}
    \item We propose a methodology to evaluate the robustness of audio embeddings against channel effects, both qualitatively and quantitatively, in a task-free setting. 
    \item We investigate the effectiveness and limitations of correlating three distance measures quantifying change or shift in pairwise distances, topology, and distribution with the downstream performance.
    \item Mimicking channel effects with four fundamental perturbations: high pass (HP) and low pass (LP) filtering, gain, and reverberation, we show that embeddings are more robust to changes in gain and reverberation than in HP and LP.
    \item HEAR shows OpenL$^3$ to perform better than YAMNet. We get a similar conclusion, but with a closer inspection of the performance differences in each of the four perturbations.
\end{enumerate}

\begin{figure}[t!]
    \centering
    \includegraphics[width=0.97\linewidth]{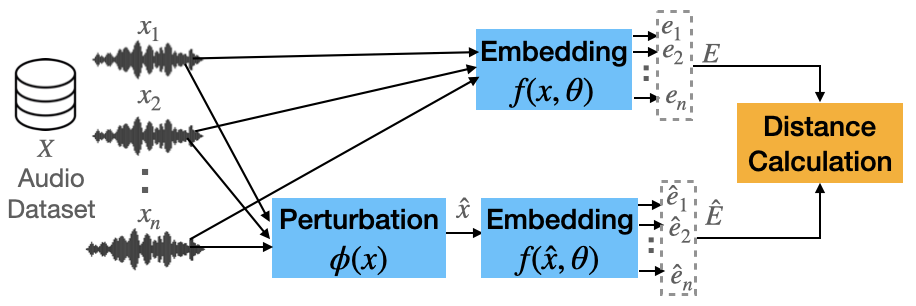}
\caption{Pipeline to evaluate the robustness of embeddings by calculating distance between the original and the new audio embeddings, $E$ and $\hat{E}$, respectively.}
    \label{fig:pipeline}
\end{figure}

\begin{figure*}[ht!]
\centering
\includegraphics[width=0.9\linewidth]{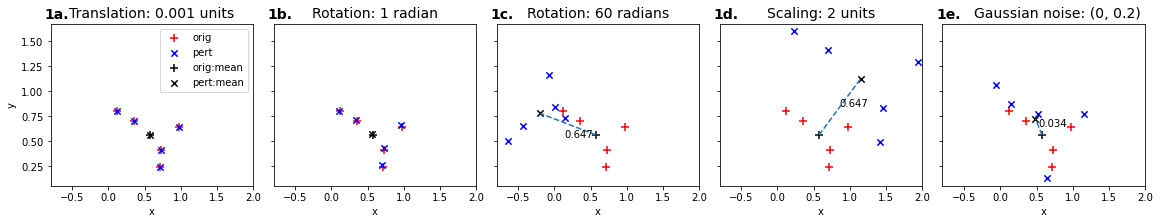}
\includegraphics[width=0.9\linewidth]{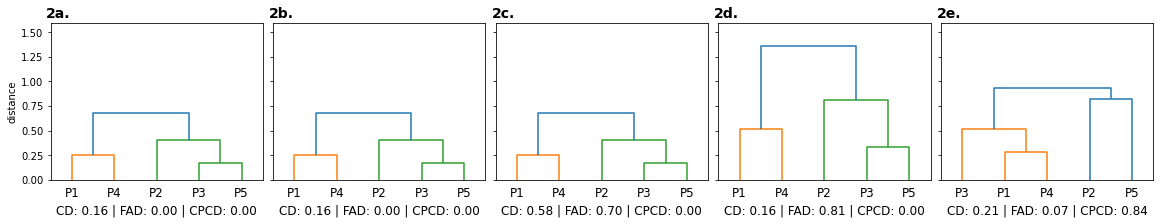}
\caption{Cosine Distance (CD), Fr\'echet Audio Distance (FAD) and Cophenetic Correlation Distance (CPCD) to measure change in pairwise distance, distribution and topology respectively for a toy dataset of five coordinates. The text on the dotted blue line denotes the distance between the original (orig) and the perturbed (pert) mean.}
\label{fig:metrics}
\end{figure*}


\section{Robustness to Perturbation}
Let $X$ = $\{x_i\}_{i=1}^n$ be a dataset of $n$ audio snippets and $\theta$ be the parameters of the upstream embedding model $f(x, \theta) \rightarrow e_x$ that maps audio input $x$ to a $d$-dimensional embedding $e_x \in \mathbb{R}^d$. $E$ = $\{e_i\}_{i=1}^n$ is the set of all such $n$ embeddings. Also, consider a transformation function $\phi(x) \rightarrow \hat{x}$ that perturbs the audio signal $x$ to $\hat{x}$. The new audio set $\hat{X}$ then produces a new embedding space $\hat{E}$ = $\{\hat{e}_i\}_{i=1}^n$. The robustness problem is then stated as follows: the embedding space $\hat{E}$ produced by the upstream model on the perturbed audio set should not change the semantics of the audio signal i.e. 
distance between $E$ and $\hat{E}$ is small. We list different distance metrics to measure the variation between the two embedding spaces in section \ref{metrics}.

We investigate four perturbations, namely \textit{high pass} and \textit{low pass} filtering, \textit{gain} and \textit{reverberation}. These perturbations are inspired by channel effects that arise when deploying environmental audio sensing devices, and simulate varying conditions both in the acoustic propagation from the source to the recording device and in the recording device itself. However, these perturbations are common in many microphone recording situations. Table \ref{tab:perturb} lists the range of values on which we explore each perturbation.

\noindent \textbf{High and Low Pass filtering}: Since low-cost microphones may not have a full frequency range response, we use high-pass and low-pass filters to approximate various frequency responses to test the representations' ability to be mic-invariant. While several applications focused on urban sound monitoring \cite{ardouin2018innovative} use MEMS mics with a frequency range of 20-20k Hz, the sampling frequency of 44.1 or 48 kHz is an expensive option for low-power micro-controllers. Recently, Lopez et al. \cite{lopez2020digital} leverage mics with a frequency range of 63-8k Hz instead. Besides the inherent differences in a mic's design, external factors like water, wind, and dust can change the frequency response. For example, water clogged inside the mic windscreen can attenuate the signal, especially at higher frequencies \cite{ribeiro2014uncertainties} and low pass filters can also simulate this.

\noindent \textbf{Reverberation}: Typically, the sources of environmental sounds are outdoors, for example, construction noise, honking, and aircraft, to name a few. However, people hearing these sounds can be in outdoor areas like streets with many buildings or indoor areas with walls and furniture. Sound waves reflect from such obstacles several times before reaching the ear. The sound reflections mix to create what is known as reverberation. We evaluate the representations for different listener environments by modeling the reverberation time of space, defined as the time required for the sound level to decay by 60 dB after the signal has stopped.

\noindent \textbf{Gain}: Due to infrastructure requirements, it is common to place the microphones far from the sound sources when collecting environmental sounds. For a spherical wave, the sound pressure level (SPL) decreases by 6 dB (one-half) per doubling
of distance from the source. For line sources such as traffic noise, the decay rate varies between 3 and 4 dB \cite{kinsler2000fundamentals}. In order to test the near-field performance of the learned representations, we vary the gain of the signal. 


\begin{table}[htbp]
\centering
\caption{Range of values for each perturbation (pert.) type for OpenL$^3$ and YAMNet}
\begin{tabular}{c|l}
\hline
\textbf{Pert. Type} & \multicolumn{1}{c}{\textbf{Pert. Values}} \\ \hline
High Pass & \{100, 200, 400, 800, 1600, 4k\} Hz \\ \hline
Low Pass & \{8k, 4k, 1600, 800, 400\} Hz \\ \hline
Reverberation & \{25, 50, 75, 100\} \%
 \\  \hline
Gain & \{3, 6, 10, 20, 30\} dB
 \\  \hline
\end{tabular}
\label{tab:perturb}
\end{table}

\section{Experimental Design}

Fig. \ref{fig:pipeline} shows the pipeline to calculate a distance measure to quantify the effect of a perturbation $\phi$ on an embedding space $E$. 

\subsection{Distance Metrics to Evaluate Robustness}
\label{metrics}

We utilize a toy dataset of five random coordinates in Fig. \ref{fig:metrics} to motivate the usage of different distance measures. The dataset is intended to be simple and illustrative rather than directly related to the perturbations in this paper.
Even minor perturbations, such as those in $1a$ and $1b$, change the pairwise distances between the old and new points. However, the distance between them remains preserved in the new space, which is evident from the hierarchical clustering in $2a$ and $2b$. Similarly, scaling by a constant factor of 2 clusters the new points the same as that in the original dataset but changes the mean of the new distribution. Motivated by the unique qualitative and quantitative information provided by various metrics, we investigate distance measures to evaluate the shift between the original and the perturbed embeddings in three aspects: (1) pairwise distances, (2) relative pairwise distances (as in hierarchical clustering topology), and (3) distribution.

\noindent \textbf{Pairwise:} When comparing embeddings, a common method has been to use some pairwise distance. We choose \textit{cosine similarity} because it is common to normalize embeddings before training the downstream classifier. To change similarity into distance, we use \textit{cosine distance} (CD). To generate a single distance value for the full embedding set, we find the mean of all the CDs. 


\noindent \textbf{Topology:} As shown in Fig. \ref{fig:metrics} $2a$-$2d$, there may be scenarios in which pairwise distances might significantly change, even when the relative distances between the data points do not vary as observed in clustering. In such situations, classes may still be well-separated in the embedding spaces, but new data may be required to represent those class distributions. To make the pairwise study less stringent and distance-free, we evaluate the total change in the pairwise proximity of the embeddings in $E$ and $\hat{E}$. We use agglomerative clustering with Euclidean distance and average linkage criterion to create dendrograms for the original and perturbed embeddings. The branching patterns (also known as topology) in the two dendrograms might differ in terms of the embedding positions in the leaf set. To quantify the difference, we calculate the \textit{Pearson correlation coefficient (PCC)} (Equation \ref{eq:pcc}) between the cophenetic distance matrices, $C_o$ and $C_p$, for the dendrograms corresponding to the original and perturbed embeddings. We utilize Equation \ref{eq:cpcc} to convert the correlation into a distance metric, which we refer to as \textit{cophenetic correlation distance (CPCD)}.
\begin{equation}
\label{eq:pcc}
    PCC = \frac{cov(C_o, C_p)}{\sqrt{var(C_o) var(C_p)}}
\end{equation}
\begin{equation}
\label{eq:cpcc}
    CPCD = 1 - PCC
\end{equation}
where $cov$ and $var$ correspond to covariance and variance, respectively.

\noindent \textbf{Distribution:} In order to get the variation in the distribution within the embedding space, we leverage the \textit{Fr\'echet Audio Distance (FAD)}. Initially proposed for music enhancement application, Kilgour et al. \cite{kilgour2019frechet} use FAD to compare the embedding statistics generated on a large reference set of clean music with the embedding statistics generated on an evaluation set of enhanced noisy signals. In this work, we use FAD to compare the statistics between the original and the perturbed embedding set. 
The Fr\'echet distance (also known as Wasserstein-2 distance) between the Gaussian of the original embeddings $\mathcal{N}_o(\mu_o, \Sigma_o)$ and the perturbed embeddings $\mathcal{N}_p(\mu_p, \Sigma_p)$ is then computed as:
\begin{equation}
\label{eq:fad}
	FAD(\mathcal{N}_o, \mathcal{N}_p) = ||\mu_o - \mu_p||_2^2 + tr(\Sigma_o + \Sigma_p - 2\sqrt{\Sigma_o \Sigma_p}) 
\end{equation}
where $\mu$ represent the mean, $\Sigma$ the covariance matrix, and $tr$ the trace. Unlike the cosine and the correlation distance, FAD is oblivious to the way the embeddings are related to one other, as illustrated in $1c$ and $1d$ in Fig. \ref{fig:metrics}. It is primarily used to investigate the change in the overall embedding distribution.

\subsection{Datasets}
\label{subset}
We study the robustness of both \textit{OpenL$^3$} and \textit{YAMNet} for two popular datasets, namely \textit{UrbanSound8K (US8K)} \cite{salamon2014dataset}, and \textit{SONYC Urban Sound Tagging (UST)} \cite{cartwright2020sonyc}. These datasets complement those used in the HEAR challenge for environmental sound. For the UST dataset, we use all the 1380 recordings with verified annotations in v2.3. As for the US8K samples, we use all the $\sim$8k samples. 
To simplify the analysis, we sample one embedding per clip, which we select by computing the sound pressure levels (SPL) and retrieving the embedding where the SPL is highest. The assumption behind this is that the highest SPL level correlates with the presence of a labeled sound source.

\begin{figure*}[tb]
\centering
\begin{subfigure}{0.32\textwidth}
  \centering
    \includegraphics[width=0.98\linewidth]{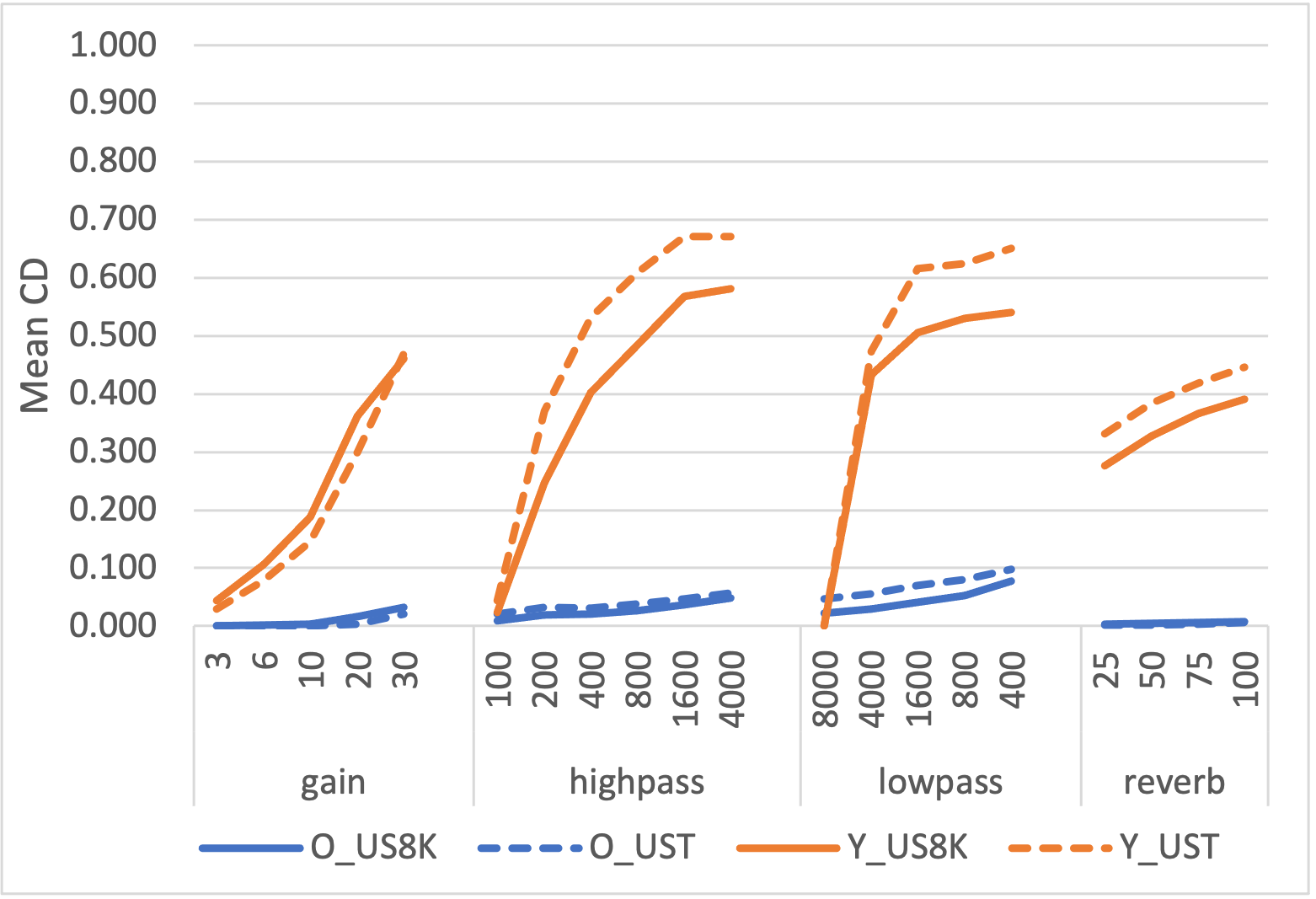}
    \caption{}
    \label{fig:cd}
\end{subfigure}
\begin{subfigure}{0.32\textwidth}
  \centering
    \includegraphics[width=0.98\linewidth]{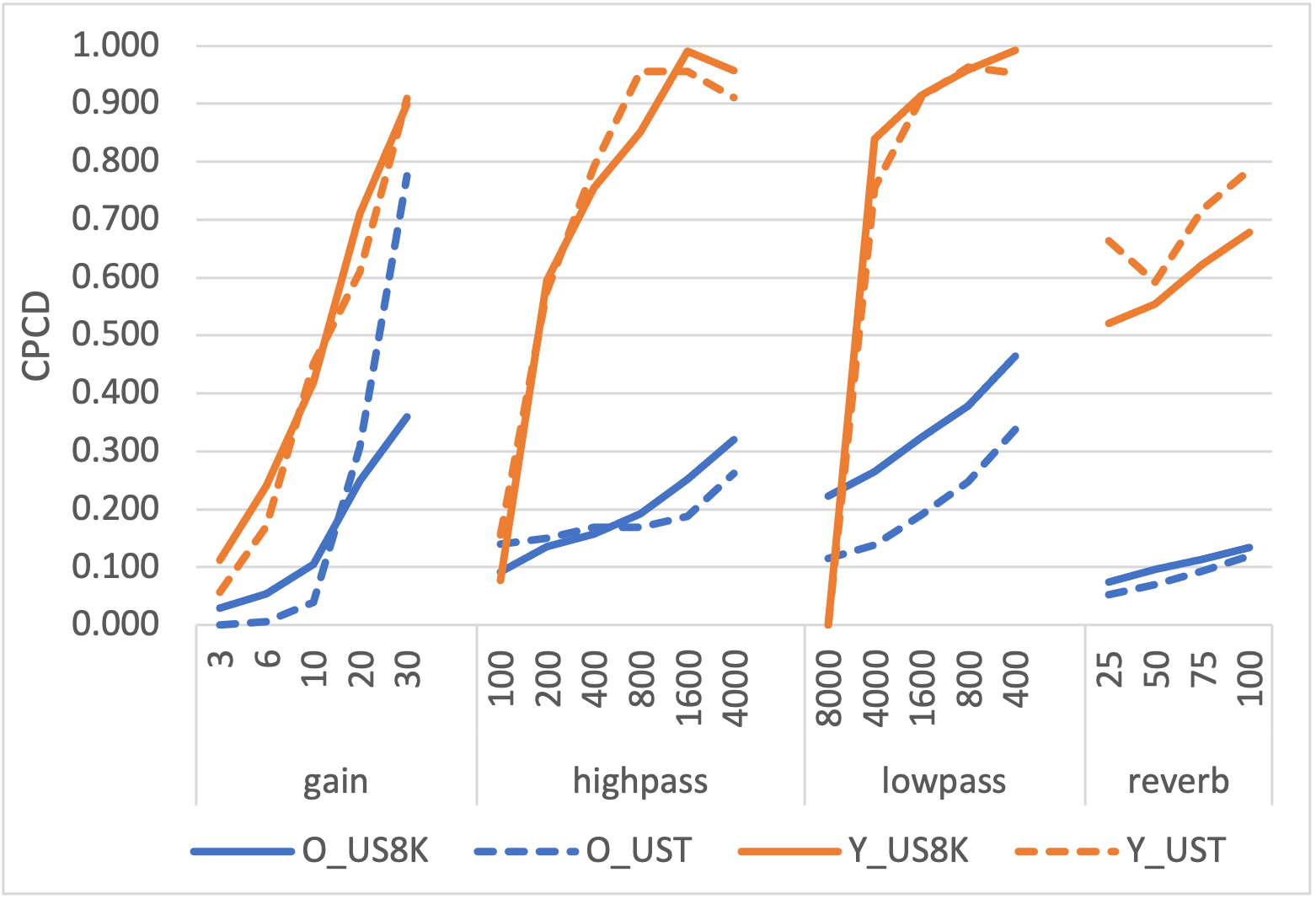}
    \caption{}
    \label{fig:cpcd}
\end{subfigure}
\begin{subfigure}{0.32\textwidth}
  \centering
    \includegraphics[width=0.98\linewidth]{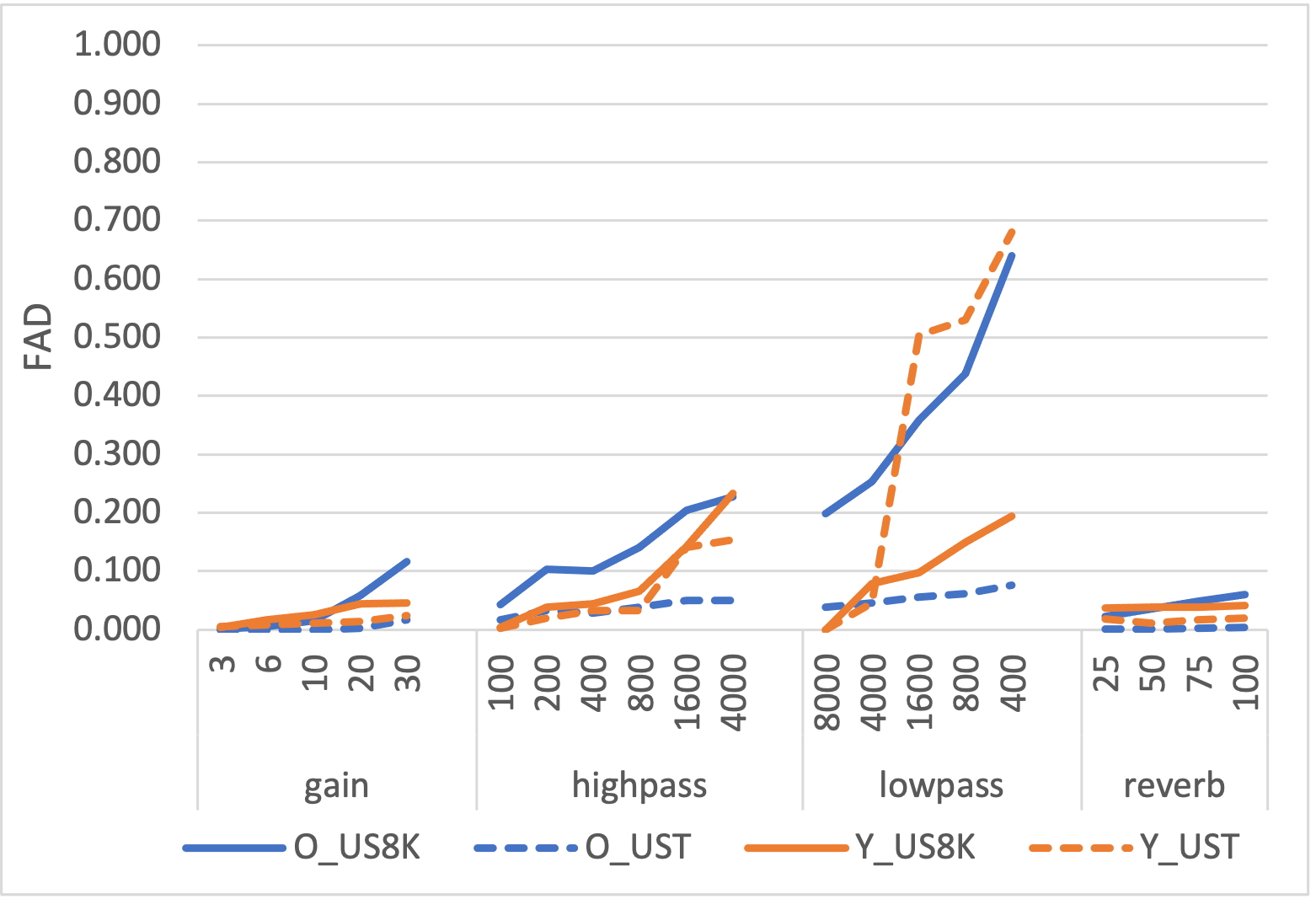}
    \caption{}
    \label{fig:fad}
\end{subfigure}
\caption{CD, CPCD, and FAD for US8K and UST datasets for four perturbation types. The x-axis values for each perturbation range from low to high changes. The sampling frequency of YAMNet (Y) and OpenL$^3$ (O) is 16kHz and 44.1kHz, respectively.}
\label{fig:eval_metrics}
\end{figure*}

\subsection{Metrics for Downstream Evaluation}
We assess the effect of perturbation not only on the embeddings but also on the classification metrics on the downstream datasets, i.e. US8K and UST. Specifically, we train a logistic regression model with the original embeddings $E$ and evaluate its performance on embeddings $\hat{E}$ perturbed with different types and values.
\\
\noindent \textbf{US8K:} We employ cross-validation accuracy as well as \textit{mean silhouette score} to compare the quality of classification and clustering of the embeddings before and after the perturbation.
\\
\noindent \textbf{UST:} As for the UST dataset, we use macro-averaged areas under the precision-recall curve (macro-AUPRC) as the primary evaluation metric. We do not use silhouette analysis for UST because it is a multi-label dataset and one embedding can be part of multiple classes at the same time.

\section{Evaluation}

\begin{figure}[tb]
\begin{subfigure}{0.48\textwidth}
 \centering
    \includegraphics[width=0.9\linewidth]{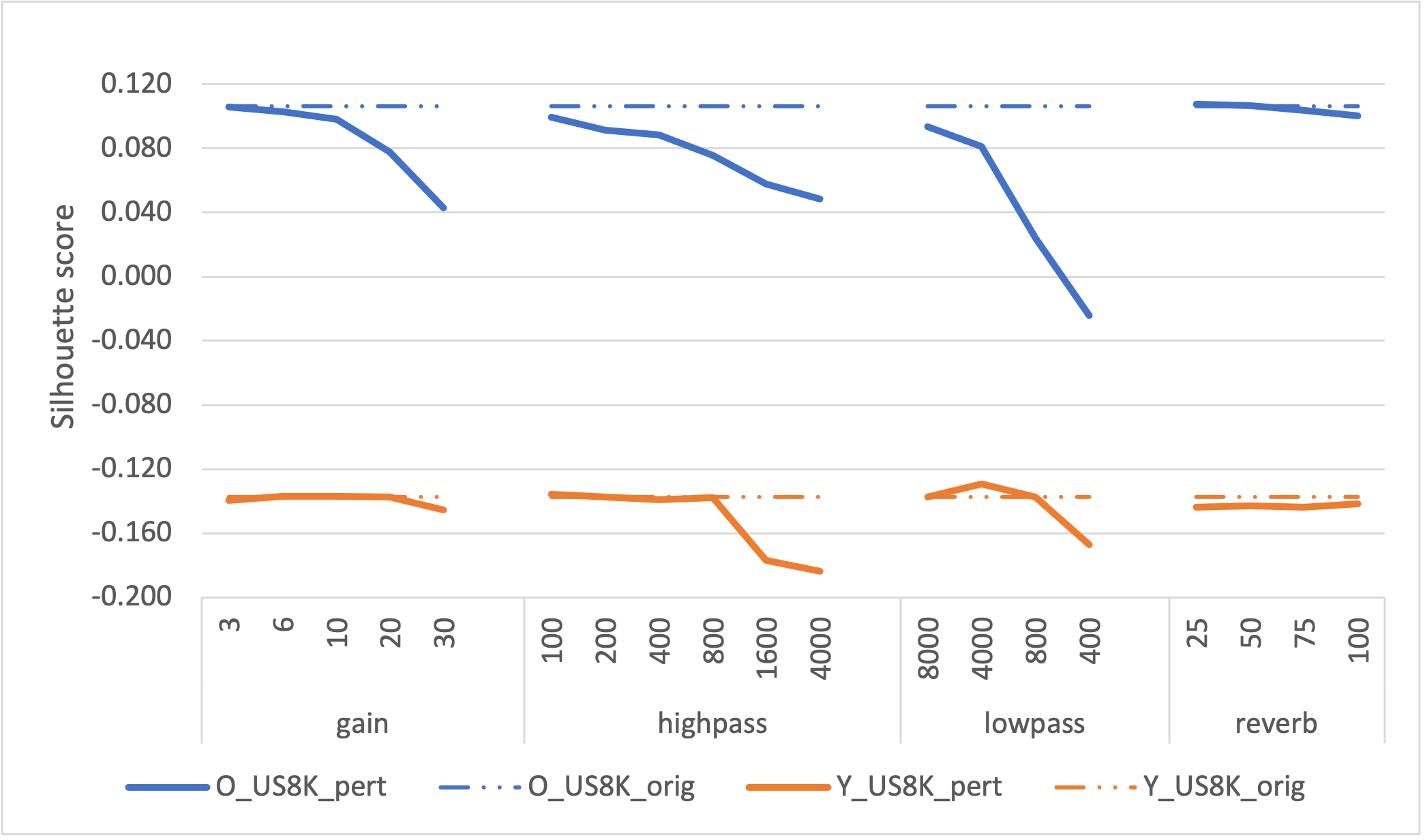}
    \caption{}
    \label{fig:sil}
\end{subfigure}
\begin{subfigure}{0.48\textwidth}
  \centering
    \includegraphics[width=0.9\linewidth]{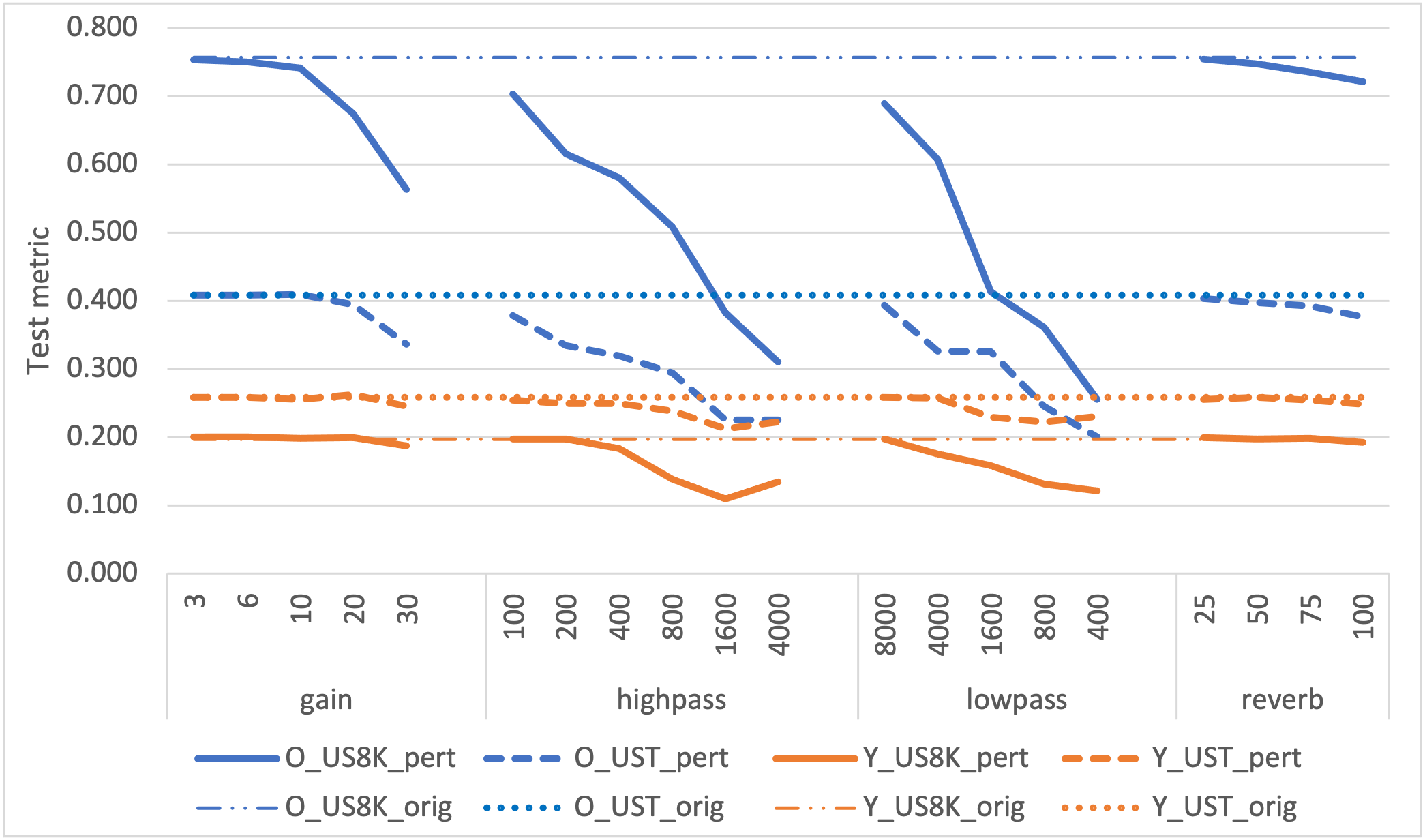}
    \caption{}
    \label{fig:accuracy}
\end{subfigure}
\caption{(a) Silhouette scores of OpenL$^3$ and YAMNet for US8K. (b) compares the classification accuracy of the original (orig) and perturbed (pert) embeddings for US8K and UST.}
\end{figure}

\subsection{Comparison of representation types}
\label{sec:compare_embs}
Looking at Fig. \ref{fig:eval_metrics}, for both CD and CPCD, YAMNet shows a larger distance (higher sensitivity) as compared to OpenL$^3$. To get a deeper understanding of YAMNet's sensitivity to pairwise relations, we calculate the \textit{silhouette} scores of the embeddings of each class in US8K (c.f. Section \ref{sec:metrics_downstream}).

Although OpenL$^3$'s distribution show more variation than YAMNet for US8K (c.f. Fig. \ref{fig:fad}) when perturbed, large values of CPCD for YAMNet in Fig. \ref{fig:cpcd} indicate that the YAMNet's pairwise relationships change significantly in the perturbed space, and can possibly affect the downstream performance. Fig. \ref{fig:accuracy} confirms this hypothesis. Note that in order to re-scale FAD to a 0-1 scale, we use Min-Max scaling within a dataset to normalize FAD scores, which somewhat skews the comparison but has no effect on the overall trend.

\subsection{Distance metrics and downstream evaluation}
\label{sec:metrics_downstream}

We compare the trends of the distance measures with the downstream evaluation metrics. In Fig. \ref{fig:sil}, we observe that YAMNet produces a negative silhouette score of $-0.14$ even for the original representations, meaning that embeddings of the same class lack the two qualities of separability from embeddings of other classes and cluster compactness. Even a tiny modification can change the pairwise groups in this scenario. This is also reflected in Fig. \ref{fig:cd} and Fig. \ref{fig:cpcd}.

The trends in FAD closely approximate the performance drops in Fig. \ref{fig:accuracy} as the severity of the perturbation increases. We notice that OpenL$^3$ has a steeper change in accuracy than YAMNet. Nevertheless, even when OpenL$^3$ produces the lowest accuracy (high pass at 8k Hz), it is better than YAMNet. One can infer the same by considering both FAD and CPCD simultaneously, as stated in Section \ref{sec:compare_embs}. The mean CD might have a neutralizing effect. For an example, let us consider two embeddings, $e_1$ and $e_2$ in $E$. If the CD decreases for the $(e_1, \hat e_1)$ pair by 0.3 but increases for $(e_2, \hat e_2)$ by 0.3, the mean change would remain unaffected. 
Both CD and CPCD utilize pairwise information and are comparatively more sensitive to noise and outliers. We recommend always supplementing such pairwise metrics with information from other robust metrics like FAD. 

Furthermore, because FAD best reflects the performance loss, it may be used for data augmentation to make the downstream classifier more robust. When it comes to determining what values to utilize for augmentation, we can see from FAD (c.f. Fig. \ref{fig:fad}) that each embedding and dataset combination appears to have different inflection points, i.e., where changes in distance and performance drop more dramatically. We believe that this value is a useful indicator of how much perturbation to use for the augmentation, as larger values may be associated with dramatic changes in the signal, while smaller values may not make a significant difference. In future research, we'll investigate the use of inflection points for augmentation.


\subsection{Comparison of perturbation types}

The embeddings are more robust to \textit{gain} and \textit{reverb} than to high and low pass filtering. This is not surprising because these perturbations do not significantly change the information contained in the signal (much less than low and high pass filtering), so the fact that the embeddings are robust to them is a good indication that the models are doing what we expect and they are mainly learning semantic information. The inflection point for FAD and CPCD at a gain of 10 dB indicates the presence of harmonic distortions associated with clipping. It is a bit surprising how much OpenL$^3$ embeddings change in response to low pass perturbations. We hypothesize this is due to a codec-related ``shortcut'' \cite{doersch2015unsupervised} in the self-supervised audio-visual correspondence task in which the model finds a relationship between high-frequency absence and image artifacts in low-bit-rate encodings.

\section{Conclusion}
We employ three distance metrics to estimate the effect of channel effects on two representations, OpenL$^3$ and YAMNet. We demonstrate that the downstream performance and the distance measures are complementary. Limiting the evaluation to downstream performance precludes a more in-depth study of the reason and extrapolation of the findings to other real-world test scenarios. Similarly, the analysis of distance measurements can be misleading when using only one metric. In our study, the combination of FAD and CPCD gave the most valuable insight and was representative of downstream trends. We recommend using FAD to choose among different perturbations for augmentation to make sound event detection models more robust. 

In future work, we intend to repeat this study on a wide variety of embeddings and datasets and extend the analysis to include correlations between distance metrics and different sound event classes.



\bibliographystyle{IEEEtran}
\bibliography{main}
\end{document}